# An application of an Embedded Model Estimator to a synthetic non-stationary reservoir model with multiple secondary variables

**Colin Daly**

Schlumberger Ltd. 55 Western Avenue, Milton Park, Abingdon, OX14 4RU

**Correspondence:**
Colin Daly
cdaly@slb.com

**Abstract**

A method (Ember) for non-stationary spatial modelling with multiple secondary variables by combining Geostatistics with Random Forests is applied to a three-dimensional Reservoir Model. It extends the Random Forest method to an interpolation algorithm retaining similar consistency properties to both Geostatistical algorithms and Random Forests. It allows embedding of simpler interpolation algorithms into the process, combining them through the Random Forest training process. The algorithm estimates a conditional distribution at each target location. The family of such distributions is called the model envelope. An algorithm to produce stochastic simulations from the envelope is demonstrated. This algorithm allows the influence of the secondary variables as well as the variability of the result to vary by location in the simulation.

## 1    Introduction

An issue that practitioners of spatial modelling confront is that the ensemble behavior of the modelled random function might not match observable properties of the data set in some types of application. For example, some auxiliary hypothesis of a generative model such as stationarity, needed to allow inference, may leave an unwelcome trace in the predicted results. The subsurface reality itself is rarely stationary and the usual remedy for scientists is to subdivide the domain of interest into regions which they believe to be approximately stationary (or reducible to stationary by removing a trend) and to create separate models within each region. While this generally works well in practice, it does suffer from a couple of potential downsides. Sometimes the number of regions required can be quite large, necessitating a labor intensive and quite error prone procedure. Secondly, a region which is modelled as stationary may in fact have some subtle, but distinct sub-regions which, if identified, would affect some desired applications of the model (for example fluid flow strongly depends on vertical spatial divisions).

A compounding factor is that, when predicting a target variable at a location, there often are a number of secondary variables known at that location which are covariate with the target variable of interest. An example from the oil industry is prediction of porosity away from well locations, where the porosity value is known to a reasonably high degree. At target locations, we may have observed one or more seismic, and/or geological attributes, which are covariate with porosity. These are known as secondary variables in Geostatistics. An observable spatial trend when seen in the data is often modelled as a polynomial. In many generative models, such as Gaussian Cosimulation, a simplified relationship between the covariates and the target variable is assumed. For example, stationarity of the correlation between secondary variables and the target variable is an auxiliary hypothesis that is unlikely to be fully satisfied in practice.

In this article, a simple alternative procedure is proposed which is aimed at reducing these effects. Influenced by the idea of Conditional Random Fields (CRF) and the classic Geostatistical wariness of very highly parametric models, the idea is to (partly) estimate the conditional distributions directly based on the secondary data and on prior speculative spatial models. These distributions can be quite non-stationary reflecting local heterogeneity patterns. They provide estimates of the target variable as well as a locally adapted estimate of uncertainty. This can be done without invoking a full multivariate distribution. Unfortunately, it is not possible to extend the workflow to produce realizations of a random function without such an assumption. In most traditional formulations, this is made up front. Examples are Gaussian Random Fields (GF), Markov Random Fields (MRF), Hidden Markov Models (HMM) and many variants. Since the prior model is generally made with some hypothesis of stationarity, the risk of this hypothesis persisting into the results should be considered.

For the approach considered here, the realizations are made by sampling from the distributions that we have estimated. However, it is only the sampling that needs to be assumed as stationary. Hence, the fully spatial model is only for the sampling function rather than as a direct model of the earth itself. To summarize:

a) A machine learning/conditional distribution estimation algorithm is used to get a distribution at each spatial location. This family of distributions is called the envelope in this paper.
b) A stationary ergodic random function is used to sample from the envelope at each location.

As well as being the basis for simulating realizations, the envelope of conditional distributions can be used to obtain estimates of the mean, quantiles, robust estimates of uncertainty and unbiased estimates of 'sweet spots' such as locations where porosity is expected to be higher than a user set threshold. The idea of 'embedding' prior spatial models in the estimation of the envelope is what gives rise to the name Ember, which stands for embedded model estimator. This allows use of an essentially unlimited number of additional variables to enhance conditioning of the final model. It turns out to still be possible to make inference about the type of variogram to use for the stationary sampling RF depending on the assumptions about the type of sampling that we wish to use. A major advantage of this method is that the final estimates are now allowed to be non-stationary. In other words, the predictor may 're-weight' the importance of the variables in different parts of the field to adapt to local heterogeneity patterns.

After a more technical introduction to the method, an application to a synthetic subsurface reservoir model is shown. Additional technical discussion of the algorithm is held back to the appendix. A more technically focused paper, and some alternative examples are in Daly, 2020.

## 2     Embedded Model Decision Forests

The envelope in Ember is the set of distributions of the target variable at the set of target locations. It can be thought of as a generalization of the trends that are used in classical geostatistical models. In the classical model, the trend is a point property (i.e. there is one value at each target location) and is often considered to be an estimate of the mean or of the low frequency component of the model. It is not exact, in the sense that it does not honor the observed values of the target variable. Typically, it is constructed either as a local polynomial of the spatial co-ordinates (universal kriging) or using some additional variable(s) (e.g. external drift). In the Ember model, a conditional distribution is estimated at each location. In an analogy with the trend, the conditional distribution is built using the spatial co-ordinates and additional variables. In addition, the envelope estimation step will often use the predictions of simpler models to aid calculation of the conditional distributions at each location by embedding. In this paper the embedded model is kriging. The rationale is that the secondary variables, which might include seismic attributes, stress estimates, distance to nearest fault, height above contact as well as true vertical depth, stratigraphic depth and x,y spatial locations, do not explicitly contain information about the spatial correlation which is available in kriging through the variogram. In the example we will see that including the additional information which kriging brings can help constrain the envelope. Depending on the case, it may be a weak variable, contributing little, or a very strong variable which comes close to determining the solution. We note that embedding models will take a little extra work as models do not behave exactly the same as data in training.

Conditional Random Fields (CRF), Lafferty et. al, 2001, avoid construction of the multivariate law. The advantage in direct estimation of each conditional distribution in the envelope compared to a generative Bayesian model is that no effort is expended on establishing relations between the numerous predictor variables. In a full spatial model these involve stringent hypothesis such as the stationarity of the property of interest (perhaps coupled with some simple model of trend) and the stationarity of the relationship between the target variables and the explanatory variables (e.g. the hypothesis that the relationship between porosity and seismic attributes do not change spatially). The principle impact of stationarity in the classic model is seen in stochastic realizations which need to invoke the full multivariate distribution and therefore lean heavily on the hypotheses. This can be greatly reduced in the current proposal.

The form of CRF that we use here to calculate the envelope accommodates and embeds existing spatial models using a Markovian hypothesis. Let $Z(x)$ be a target variable of interest at the location $x$, and let $\boldsymbol{Y}(x)$ be a vector of secondary or auxiliary variables observed at $x$. Let $\{Z_i, \boldsymbol{Y}_i\}$ be observations of the target and secondary variables observed in the field, i.e. $Z_i$ denotes the value of the target variable $Z(x_i)$ at training location $x_i$. Finally let $\boldsymbol{Z}_e^*(x) = \boldsymbol{f}(\{Z_i, \boldsymbol{Y}_i\})$ be a vector of pre-existing estimators of $Z(x)$. Then the Markov hypothesis that we require is that the conditional distribution of Z(x) given all available data $\hat{F}(z|\boldsymbol{Y}(x), \{Z_i, \boldsymbol{Y}_i\})$ satisfies,

$$\hat{F}(z|\boldsymbol{Y}(\boldsymbol{x}), \{Z_i, \boldsymbol{Y}_i\}) = E[\mathbb{I}_{Z(x)<z}|\boldsymbol{Y}(x), \{Z_i, \boldsymbol{Y}_i\}] = E[\mathbb{I}_{Z(x)<z}|\boldsymbol{Y}(x), \boldsymbol{Z}_e^*(x)] \quad (1)$$

This hypothesis states that the conditional distribution of $Z(x)$ given all the secondary values observed at $x$ and given all the remote observations of $\{Z_i, \boldsymbol{Y}_i\}$ can be reduced to the far simpler conditional distribution of $Z(x)$ given all the secondary values observed at $x$ and the vector of model predictions at x.

In this paper, we choose to base the algorithm for calculation of the envelope on a highly successful non-parametric paradigm, the Decision Forest e.g. Meinshausan, 2006. An introduction to the method is found in appendix 1. A decision forest is a set of decision trees. For the training stage, each tree starts with a single node containing all the training data. For now, let us ignore the role of the embedded models. The training data are therefore vectors $(Z_i, \boldsymbol{Y}_i)$. Each node is split reclusively using a threshold on one of the variables $Y_k$, for some k, in a way that helps to improve the fit until the terminal nodes contains (typically) a single sample. To predict at a location, $Z(x)$, with secondary vector $Y(x) = y$ for the single tree, the value of $y$ is 'dropped' down the tree and the prediction of Z(x) is the value of $Z_i$ found in the terminal node where $y$ ends up. Each tree in the forest is generated with some random parameters meaning that the predicted result can change from one tree to the next. The final estimator of the conditional distribution is of the form

$$\hat{F}(z|Y(x), \{Z_i, Y_i\}) = \sum_{i=1}^{n} \omega_i(y) \, \mathbb{I}_{\{Z_i < z\}} \qquad (2)$$

where the weights $\omega_i(y)$ count how frequently the value of $Z_i$ is used as predictor. Under certain conditions, it can be proved that $\hat{F}(z|Y = y) \to F(z|Y = y)$ as $n \to \infty$ (appendix 1).

Embedded models are treated slightly differently to secondary data. They are embedded (appendix 1) by training our decision forest on cross validated estimates. Thus, our training data set for each tree is $\{Z_i; Y_i, Z^*_{CVe}(x_i)\}$, where $Z^*_{CVe}(x_i)$ are cross validated model estimates at training location $x_i$. With an estimate of the conditional distribution now available at every target location, it is a simple matter to read off estimates of the mean of this distribution, which we call the Ember estimate. When kriging is the strongest variable the Ember estimate is usually close to the kriging estimate but will typically be better than kriging if the secondary variables are the strongest. It must be noted that the Ember estimate, unlike kriging, but in analogy with trend modelling, is not exact. It is also possible to quickly read off quantiles, measures of uncertainty such as spread, P90-P10, as well as interval probabilities of the form $P(a \leq Z(x) \leq b)$.

Finally, for many applications we require a means of producing conditional simulations. The idea here to model $Z(x)$ is simply to take a conditional sample from the envelope of distributions $F(Z(x)|Y(x) = y)$ using a uniform stationary, ergodic random field $U(x)$ such that the result is conditioned at the hard data locations. If we use a transformed Gaussian random field for $U(x) = G(X(x))$, for a Multigaussian random field X(x), then this can be achieved by a truncated Gaussian simulation. Moreover, in this case, we can obtain a relationship between the experimental variogram for our target variable Z(x) and that of the sampling random function allowing appropriate model fitting. An approximate version of this relationship when Z itself is close to Gaussian is

$$\rho(x_1 - x_2) = E\left[\left(\frac{Z(x_1) - \hat{\mu}(x_1|y)}{\sigma_{x_1}}\right)\left(\frac{Z(x_2) - \hat{\mu}(x_2|y)}{\sigma_{x_2}}\right)\right] \qquad (3)$$

where $\rho(h)$ is the covariance function of the Gaussian RF, X(x), to be simulated, $\hat{\mu}(x|y)$ is the Ember estimate and $\sigma_x$ is the standard deviation of the residual $Z(x) - \hat{\mu}(x|y)$ which can be read from the estimate of the conditional distribution at x.

## 3      Application

A synthetic reservoir was constructed to allow for exploration of some of the issues raised in realistic surroundings. For a real case study, see Daly et al. 2020. For a longer introduction to the current example that explains geology, model construction and some technicalities of the problem, see Daly, 2020 (arXiv). The model alternates between marine dominated stacked shoreface sands in which the porosity varies quite smoothly, prograding in the distal direction which is to the South West, and fluvial systems which vary in terms of their net to gross. Figure 1 shows the 'true' porosity in the reservoir and a Relative Acoustic Impedance (RAI) image. The true porosity is unknown and is the target variable to be estimated. We will consider two cases, one with few training data, the other with more data. These are 8 and 36 wells respectively and their positions are shown on figure 1.

Figure 2 shows the wells on a plan view of the reservoir in one of the shoreface layers. In the 8 well case, 6 of the wells are in the central fault block, with 2 wells in the block to the south west where the best porosity for shoreface sands is usually to be found. There are no wells in the smaller block to the north west. The 36 well case has a better distribution of well location. The thick red line on the images represent the location of the cross section that is used in many of the subsequent figures. In the 8 well case, there are two wells in quite close proximity to the section, one at either end, but the center is not particularly well controlled by well information.

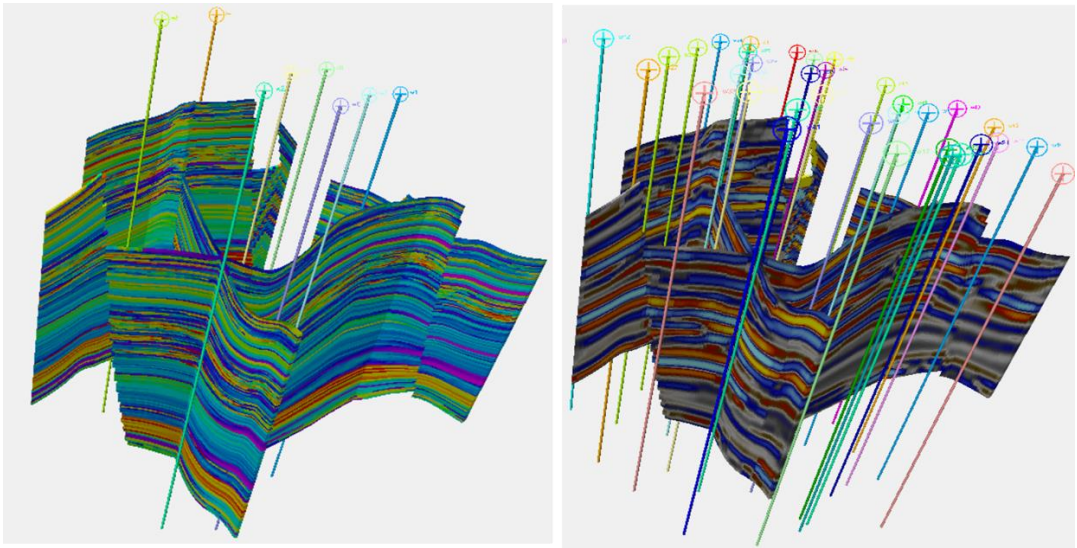

**Figure 1:** On the left, the true porosity of the reservoir to be modelled. On the right, a synthetic RAI volume created. The locations of the eight first eight wells are shown on the left, while the thirty six well case is on the right.

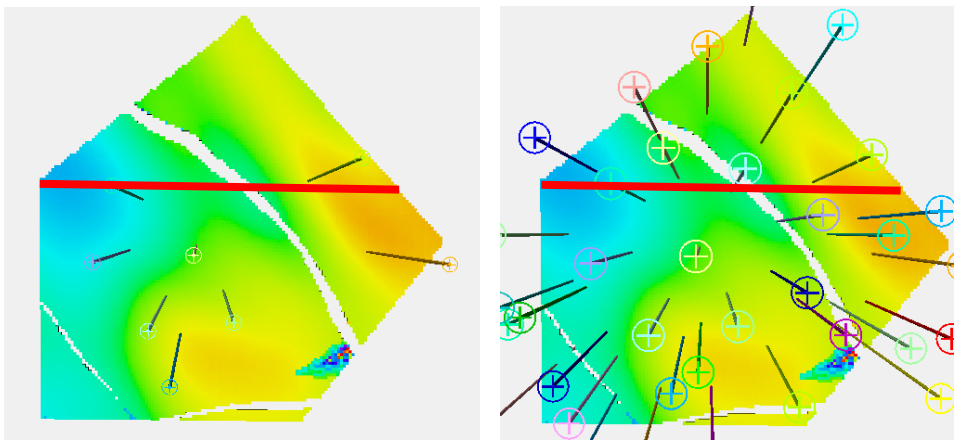

**Figure 2:** Plan view with the 36 well configuration on the right and the 8 well one on the left

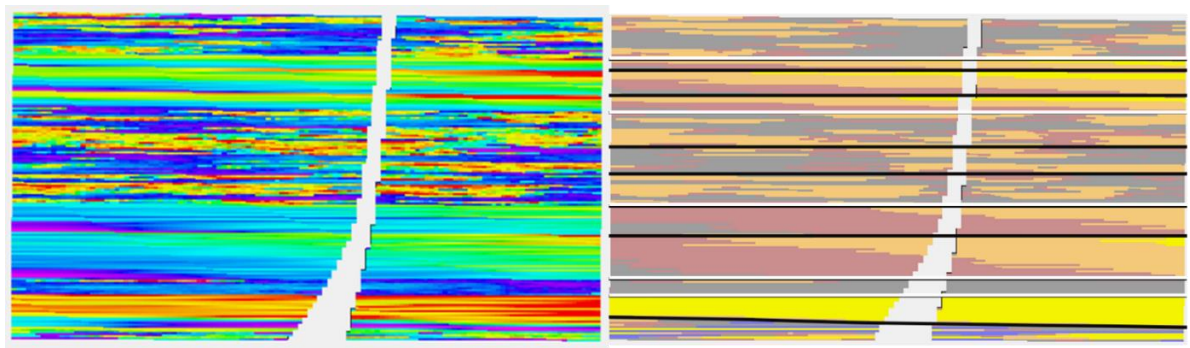

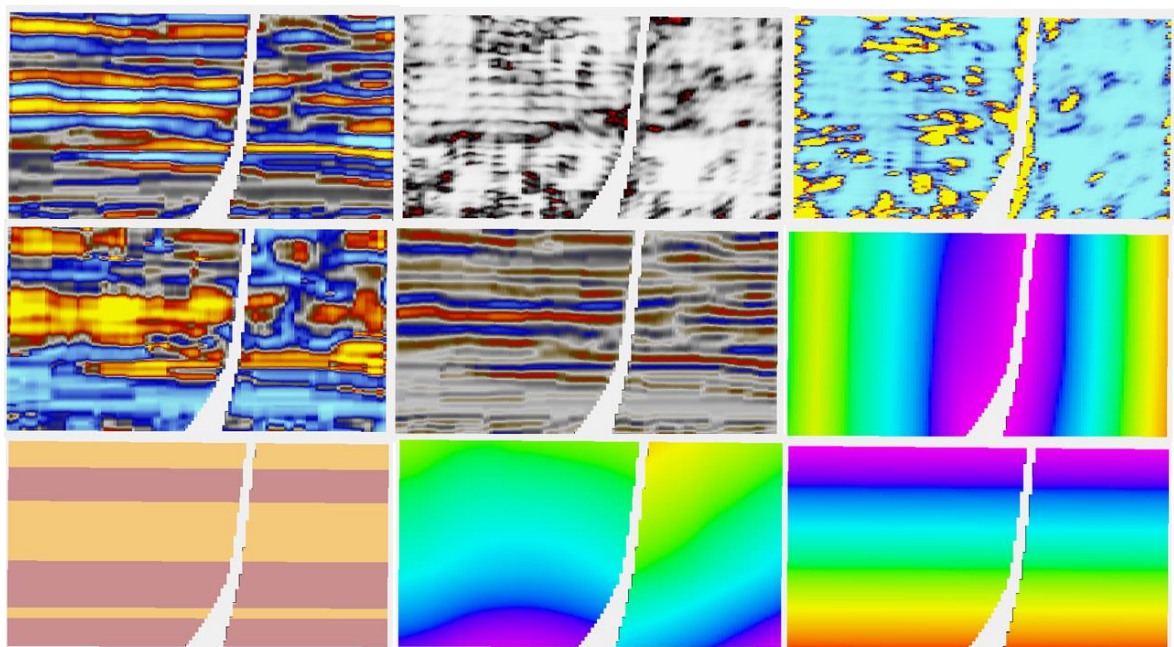

**Figure 3:** Cross section of the reservoir. The facies used for model construction are shown on the right, although these are not used in subsequent modelling effort.

**Figure 4:** Cross section of nine of thirteen data variables used in modelling. From L to R per row: RAI, Chaos, Flatness: Sweetness, Amplitude, Dist. to Fault: Depositional Zones, TVD, Strat. depth.

A synthetic seismic volume was created from a slightly modified version of the reservoir to ensure that it does not correspond too neatly to the 'real' reservoir. Several attributes were derived from this volume. The modelling was performed using 13 data variables and 2 embedded kriging models giving a total of 15 secondary variables. The 13 data variables consisted of 5 seismic variables and 8 geometric variables. Nine of the data variables are shown in figure 4.

For simulations, a variogram is required for the Sampling Random Function. In many real-world reservoir modelling cases, there is not enough information to calculate reliable variograms, especially in the horizontal direction. In such cases, the range is considered an uncertainty. The same is true using the Ember model, although as we shall see, the uncertainty is often somewhat mitigated by the estimate of the conditional distribution. In the case that a meaningful fit for the variogram of the target variable can be found for classical modelling, then using equation 3, it is likely that there is also a route to find the variogram for the Sampling RF (see Daly(2020) for an example). Indeed, for the 36 well case, it was possible to obtain a reasonable estimate for the Sampling RF's variogram. This was not the case for the 8 well situation, so, as usual, users will need to consider robustness and uncertainty of the estimation and simulation with regard to a poorly defined variogram.

Figure 5, which deals with the model of layer 91 in the case of having 8 wells, has two parts, A and B. The 4 figures on the left are part A. It is a layer with quite a low net to gross. The channels are partly identified by the seismic. In 5A the truth (i.e. the target porosity variable we are trying to estimate) and RAI as well as estimates of the spread (P90-P10) and the mean of the estimated conditional distributions are shown. With only 8 wells, the variability of the mean estimate is low compared to the true distribution. Remember that the estimated mean does not exactly honor wells, though is often fairly close. Call this the prior mean.

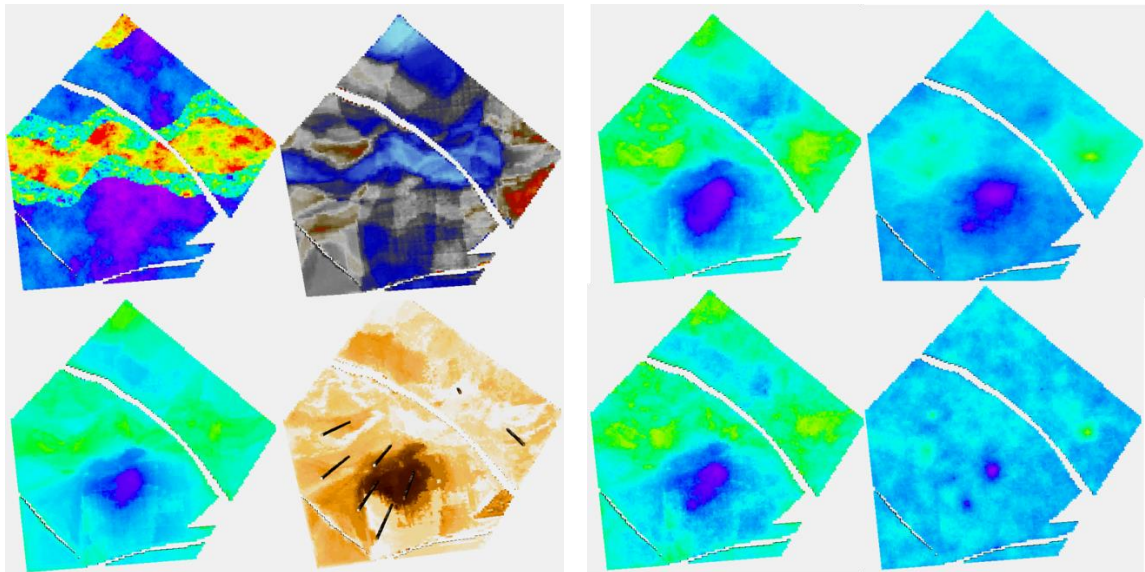

**Figure 5 A and B:** Eight well case. Figure A is on the left. Clockwise, Truth, RAI, Uncertainty Spread (P90-P10) and Prior mean; Figure B on the right. Clockwise, Posterior Ember Mean-Good variogram(G.V), Posterior Gaussian Mean-G.V, Posterior Gaussian Mean-Poor Variogram(P.V). Posterior Ember Mean-P.V

Since Ember simulations condition to wells, we calculate a posterior mean by averaging many realizations. While it doesn't seem to bring much in terms of new information compared to the prior mean, and so probably doesn't need to be routinely calculated, the results of the posterior mean for Ember as well as for Gaussian simulation (sometimes called the E-type) is shown in figure 5B. To show the effect of using an incorrect variogram, the calculation is done twice, once using the exhaustive variogram and again with a model fitted to the empirical variogram. The variogram in the second case is quite wrong (having only about 15% of the range of the good case. Of the 4 estimates in 5B, the one using the Gaussian model with the incorrect variogram is by far the worst performing due to the distribution of variability in the Gaussian case being governed directly by the choice of variogram.

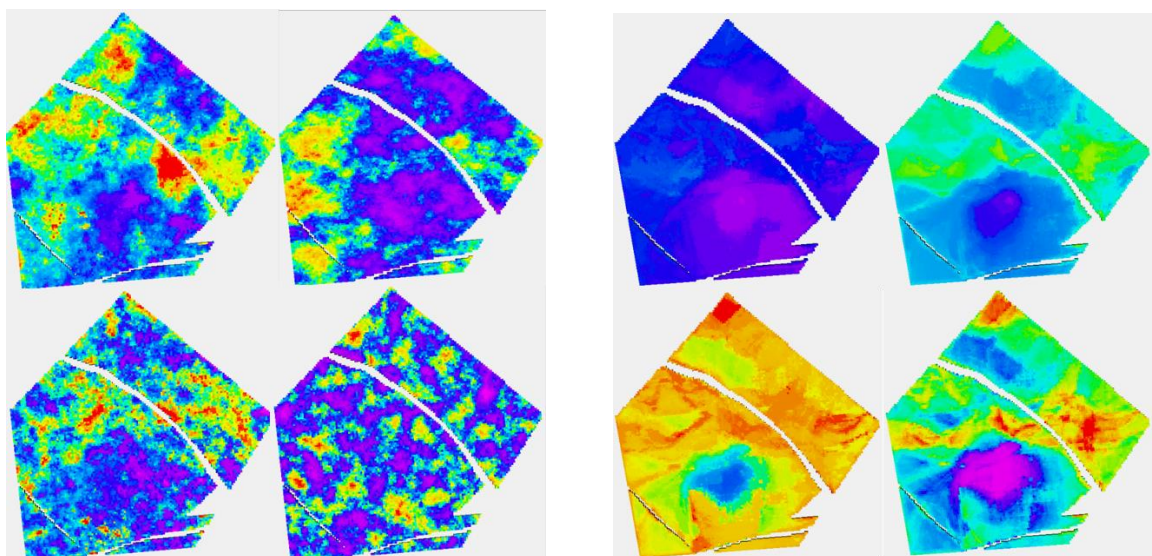

**Figure 6 A and B**: Eight well case. Left, clockwise, Ember Simulation-Good variogram(G.V), Gaussian Simulation-G.V, Gaussian Simulation-Poor Variogram(P.V). Ember Simulation-P.V. On the right, clockwise, Ember Quantiles: P10, P50, Prob(porosity>20%) and P90.

The results of stochastic simulation are shown in figure 6A. The two figures at the top correspond to a simulation from Ember and from Gaussian Simulation using the near optimal variogram and the two below using the short variogram model. Again, the relative homogeneity of the Gaussian short-range model is noticeable and is a result of the stationary hypothesis not being compatible with the true distribution, whereas the Ember result is far more robust. The figure on the right shows some by-products of the estimation phase of Ember modelling. These are the 3 quantiles, P10, P50 and P90 as well as the estimate of finding sand with high porosity of 20% or above. The true location of sand above 20% is shown in figure 7 in red.

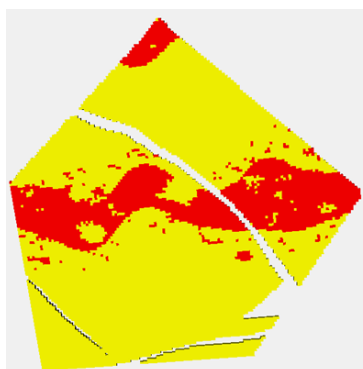

These high values lie within the channel belt. The channel belt is readily identified in the P50 and P90 cases as well as the probability map. For context, the color red in the probability map corresponds to estimated probabilities of 0.5 or above of finding sand with porosity above 20%. The spread between P10 and P90 values show that while the channel is identifiable, there is still variability and so high porosity patches are still possible outside the belt.

**Figure 7:** Red are locations when true porosity is above 20%

Turning attention to the 36 well case to get a feel for how additional information changes the Ember estimates. Figure 8 shows the same simulations and quantiles shown in figures 6 with the training using the extra wells. The improvement is quite clear in the results. The additional information identifies and isolates the belts themselves quite well but is not quite enough to determine internal heterogeneity. This is simply a function of the geology as can be seen in figure 13 which shows uncertainty of the Ember model in cross section. It is noticeable in fig 8 that the distribution of heterogeneity still depends strongly on the variogram for the classical Gaussian models even with the increased well count and gives poor results for the short variogram.

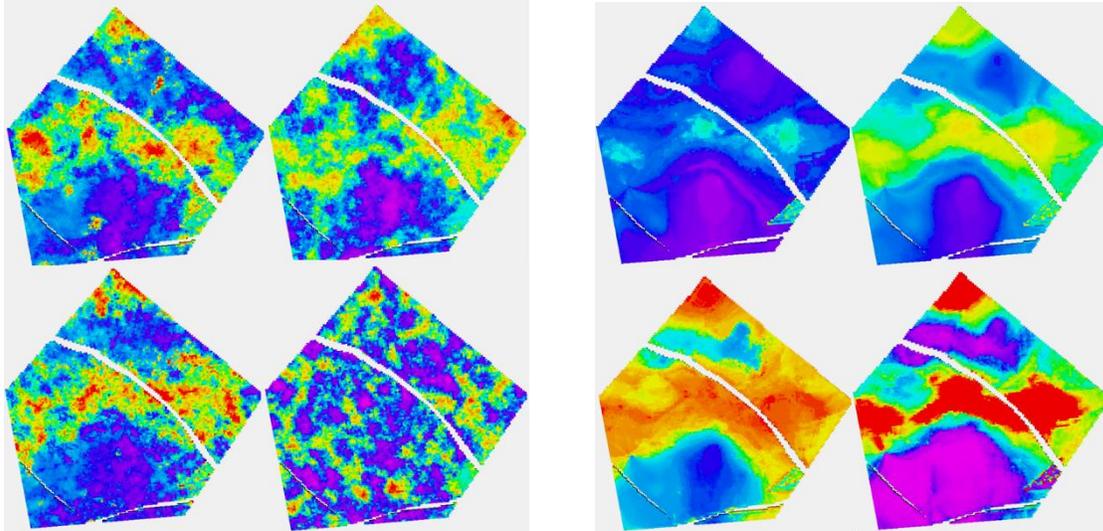

**Figure 8 A and B**: Thirty-six well case. The configuration for 8A, on the left follows that of figure 6. On the right, top row is P10, P50, bottom row is P90 and Prob(porosity>20%)

Next, consider a layer in the shoreface sands. Uncertainty is lower in the shoreface and moreover, it reduces more quickly with increasing data due to the greater simplicity. The porosity distribution is not well identified on seismic within the shoreface, so the Ember estimate is largely depending on geometric variables and the embedded kriging. Figure 9 focuses on the Ember estimation. As well at the truth and RAI, it shows estimated prior mean for the 8 and 36 well cases as well as their uncertainty estimates. Note, the patch of high porosity sand is not identified in the 8 well case but is in the 36 well case, as none of the initial 8 wells sample it. The overall trend is acceptable even with 8 wells.

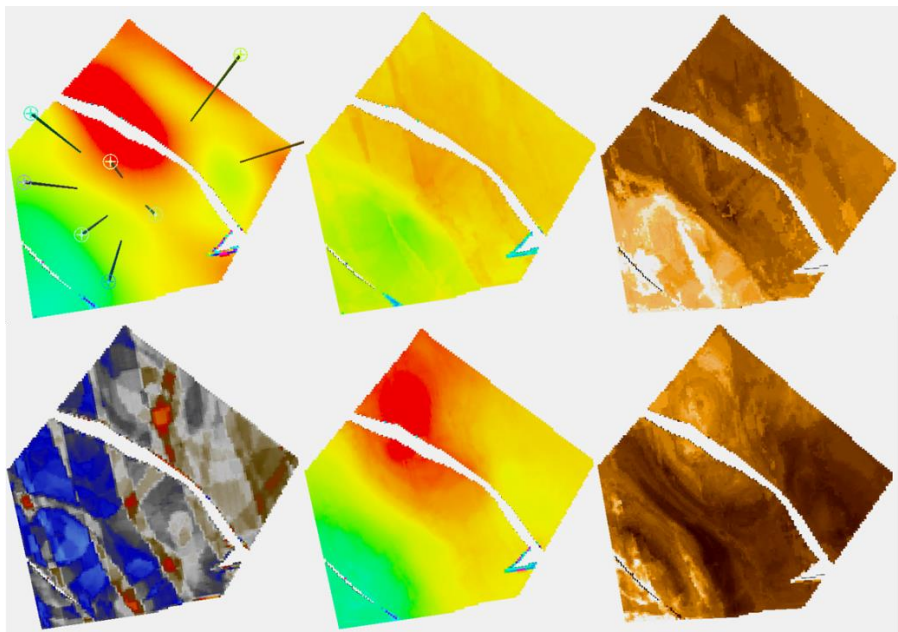

**Figure 9**: Shoreface layer. Left column, top Truth, bottom RAI: Middle column, Ember estimates, 8 well case on top: Right column, Ember uncertainty, 8 well case on top.

Ember and Gaussian simulations are shown in figure 10. As before it is noticeable that the classic Gaussian model is less robust to an incorrect choice of variogram.

In the two layers shown so far, the seismic was a large contributor for the channel case but played little role in the shoreface case. To show an intermediate situation, another layer from a channelized formation is shown in figure 11. For this layer, we just look at the Ember solution as the Gaussian one performs similarly to the previous case.

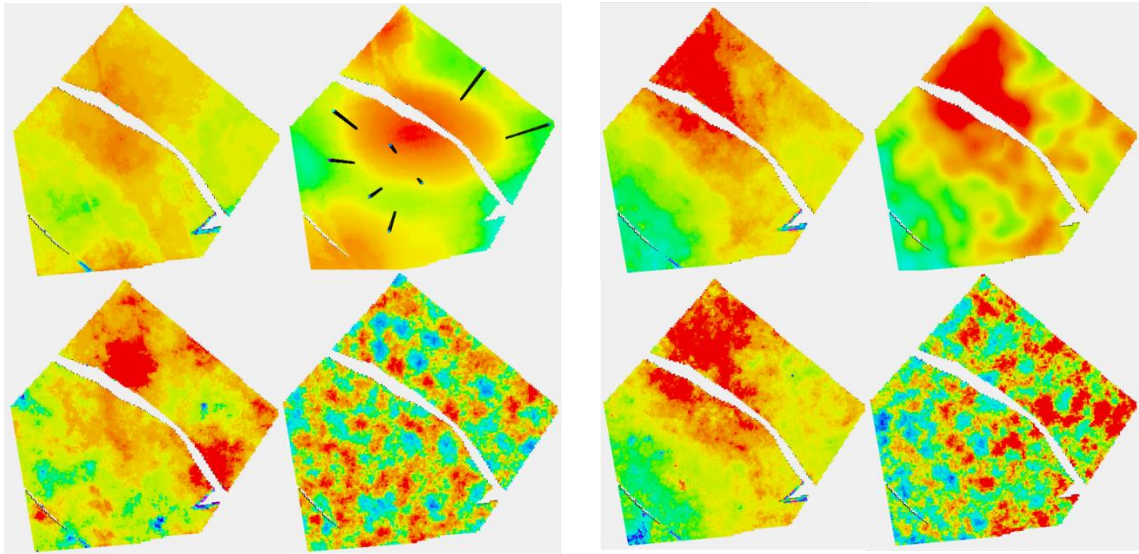

**Figure 10**: Case A, on left, is for 8 wells, B, on right, is for 36 wells. Each group is organized as before. Ember on the left, Gaussian model on right. Correct variogram on top, short variogram below

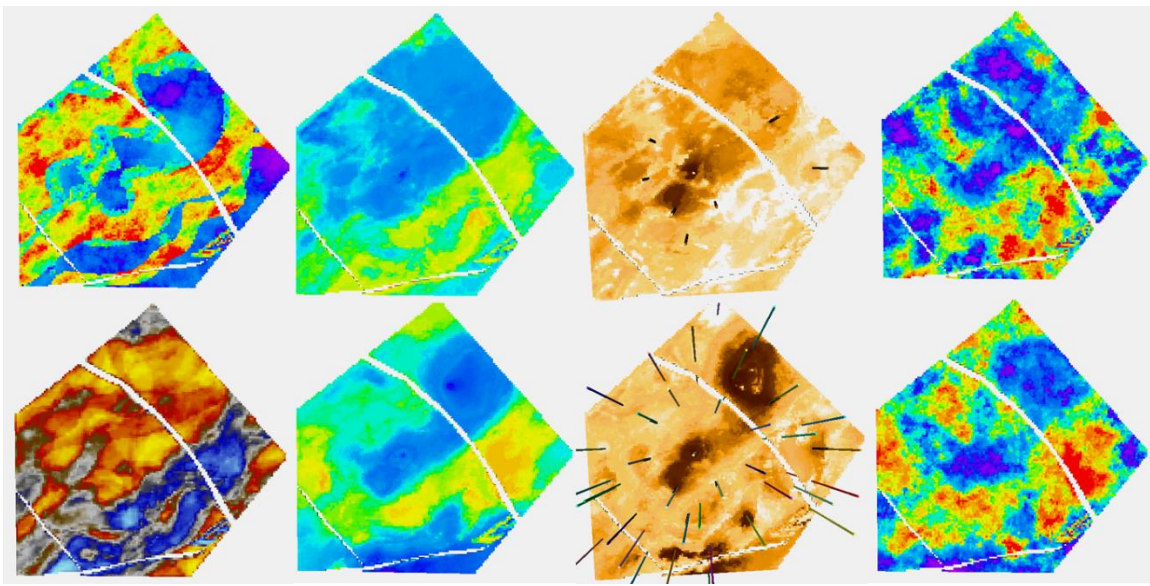

**Figure 11:** Top left is Truth, bottom left is RAI. For the other three columns, top is 8 wells, bottom 36 wells. Left to Right, mean, uncertainty and simulation from Ember algorithm.

In the top left of figure 11, two channel belts are visible, but only the easternmost one of them is readily discernible on the RAI seismic attribute. None of the 8 wells traverse the thin western channel belt, although one well goes through the overbank. Only one well goes through the thicker eastern belt. The estimate of the mean for the 8 well case fails to identify the western belt particularly to the north, but thanks to the seismic, identifies the eastern one. The same is true for the simulation. The uncertainty is sufficient to allow some sand to be located in the northern part of the western channel belt in some realizations, but it is far from systematic and with the sampling random function(SRF) being Gaussian (and hence high entropy) is unlikely to form a connected structure in any case. This simple case shows that, in some cases, it may be of interest to look at the possibility of using lower entropy SRF to provide genuinely different Ember realizations, or to use Ember to produce facies probabilities and continue with standard facies modelling methods.

In the 36 well case, both channels are well identified in the mean and the simulations tend to respect them. The eastern channel is still better identified due to the combination of seismic and embedded kriging while the western one is less defined as it is less able to exploit the seismic attribute.

So far, we have focused on plan views of several layers of the model. For completeness, and to see the value of the additional well information we return to the cross-section view in figures 12 and 13.

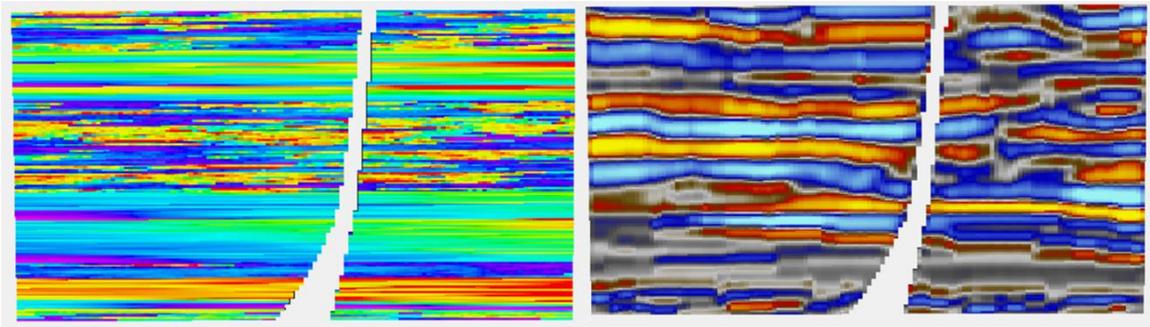

**Figure 12:** Left is True porosity while the right-hand side is RAI

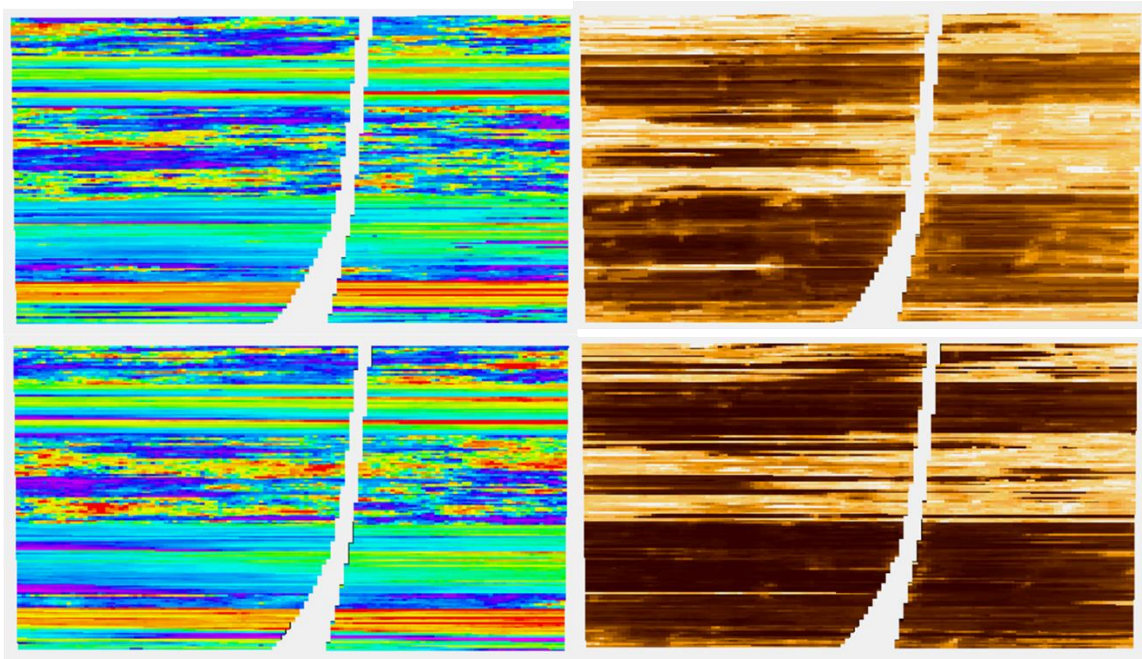

**Figure 13:** Upper images are the 8 well case, while lower are 36 well case. Left is Ember simulation while right is associated uncertainty.

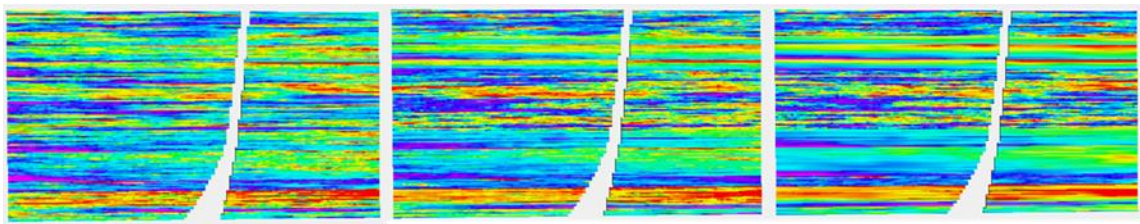

**Figure 14:** The classic Gaussian simulations referred to in Table 1. They differ by the trend management and the variogram used.

Before showing some numerical results, it is worth noting that the classic gaussian simulation model was fitted in a few different ways with separate zones, trends and covariance models. Three different results are shown in figure 14. It is not claimed that this is the best that can be done with the classic methods, just that more considerably more time was spend on these than on the Ember solution and Running Titlethey also used information that would not typically be available in real world modelling whereas the Ember model did not.

The best Gaussian model was an MM2 model with an unpublished modification developed by the author for the Petrel software suite to account for multiple secondary variables. An MM2 model really amounts to the estimation or simulation of the residuals accounting for the trends (Chiles and Delfiner, 2012). Since the residual variability is quite high and the distribution uses a stationary random function, this accounts for the fact seen earlier and again observed in Tables 1&2, that the posterior mean depends strongly on the chosen variogram so that an incorrect choice of variogram can lead to a rapid degradation of estimation and simulation quality. The major advantage of Ember simulation in this case is that it robustly distributes the errors in the right place, assuming that there is enough information to capture heterogeneity in the conditional distributions.

| 8 Well Model | Zone | Var Error | IQR Error | Var Sim Err | IQR Sim Err |
|---|---|---|---|---|---|
| **Gaussian 1** | Channel | 45.20 | 6.12 | 71.16 | 7.28 |
|  | Shoreface | 15.11 | 3.80 | 46.22 | 7.82 |
| **Gaussian 2** | Channel | 38.85 | 6.98 | 81.17 | 7.67 |
|  | Shoreface | 10.02 | 2.63 | 26.63 | 5.33 |
| **Gaussian 3** | Channel | 45.03 | 6.88 | 80.93 | 8.11 |
|  | Shoreface | 25.28 | 2.52 | 32.13 | 3.57 |
| **Ember** | Channel | 39.57 | 8.07 | 83.76 | 7.42 |
|  | Shoreface | 8.64 | 2.75 | 12.52 | 2.92 |
| **Emb Short Vario** | Channel | 39.57 | 8.07 | 88.61 | 7.62 |
|  | Shoreface | 8.64 | 2.75 | 21.64 | 3.92 |
| **Gau Short Vario** | Channel | 48.51 | 8.15 | 95.54 | 9.39 |
|  | Shoreface | 20.71 | 5.66 | 35.65 | 7.17 |

**Table 1:** Variance of Error and IQR for 3 Gaussian models and the Ember model as well as for the poor choice of 'short' variogram for both types. 8 well case. Green is best case, Red is worst case.

| 36 Well Mod. | Zone | Var Error | IQR Error | Var Sim Err | IQR Sim Err |
|---|---|---|---|---|---|
| **Gaussian 1** | Channel | 32.26 | 3.38 | 42.46 | 5.21 |
| | Shoreface | 5.39 | 1.31 | 19.13 | 4.39 |
| **Gaussian 2** | Channel | 28.28 | 3.98 | 53.38 | 6.59 |
| | Shoreface | 3.71 | 0.75 | 8.93 | 2.66 |
| **Gaussian 3** | Channel | 31.17 | 4.11 | 54.96 | 5.69 |
| | Shoreface | 6.11 | 1.43 | 7.15 | 2.12 |
| **Ember** | Channel | 26.29 | 4.35 | 52.67 | 4.51 |
| | Shoreface | 2.56 | 0.83 | 3.00 | 0.96 |
| **Emb Short Vario** | Channel | 26.29 | 4.35 | 52.67 | 4.45 |
| | Shoreface | 2.56 | 0.83 | 7.10 | 1.50 |
| **Gau Short Vario** | Channel | 43.55 | 8.21 | 92.11 | 8.92 |
| | Shoreface | 14.41 | 4.22 | 30.83 | 6.37 |

**Table 2:** Variance of Error and IQR for 3 Gaussian models and the Ember model as well as for the poor choice of 'short' variogram for both types. 36 well case

Since the Ember model estimates the conditional distribution at each location, it can be interesting to view what this envelope looks like. To do this, we have selected a location for a blind well and estimated the values there. Results of the Ember model are shown in the 8 well case in figure 15.

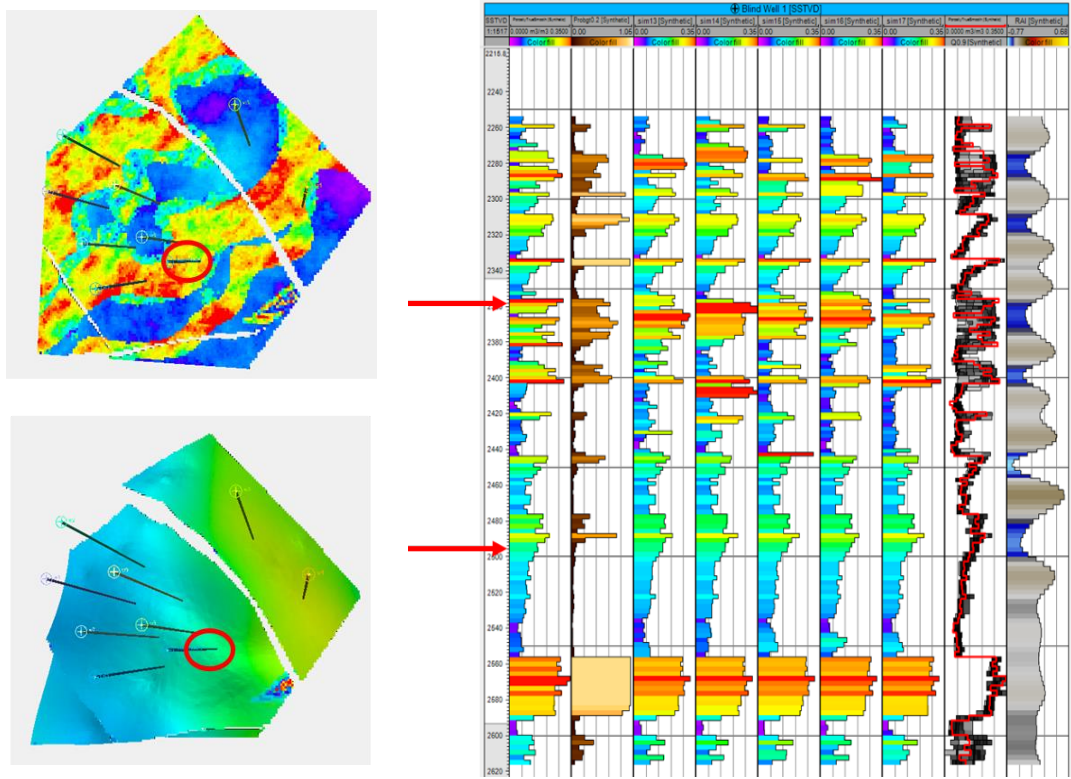

**Figure 15:** Ember results for a blind well, circled in red. Two layers are on the left, one in channel sand, the other in shoreface. Their depth locations on the well track are shown with arrows. Tracks are, from left: True porosity of blind well; Prob(poro>15%); next 5 tracks are simulations; true porosity (red) superimposed on envelope; RAI.

The results show, as expected that predictions are easier and have lower uncertainty in the shoreface sands. The first track is the true porosity at the blind well location. Each thin black line on the track is a 5% porosity increment. So, porosity values above 15% cross the 3$^{rd}$ black line. The second track shows the estimate of that the location has porosity above 15%. These probabilities are much higher, reflecting greater confidence in the estimate in the shoreface sands. So, for example there is a thin layer of shoreface sand just above the lower red arrow with porosity just above 15%. It is predicted with high probability (about 0.75) and appears in all 5 of the simulations. In fact, the simulations vary little within the shoreface and are all very similar to the true porosity log. This is confirmed in view of the envelope itself, which is the 2$^{nd}$ log from the right. It shows the true log in red, overlaying the conditional distribution which is seen as a grey scale image 'spread horizontally' at each depth. The lower bound of the 'spread' is the P10 and the upper bound the P90 of the conditional distribution. The true porosity lies between the P10 and P90 most of the time. The darker the color of the distribution at any point on the track, the higher the probability of having that value (i.e. the color represents probability density). In the shoreface sands we see that distribution is very concentrated, meaning the conditional distribution has low variability.

In the channel sands, the conditional distribution has got a far larger spread. Interestingly at some depth values, near the top for example (e.g depth=2280), the distribution starts very dark on the left, becomes paler in the middle and returns to dark on the right-hand side. This tells us that the distribution is bimodal at that depth. On reflection, this is not surprising. The prediction knows that it is in a channel interval and within those intervals it must either be in channel or not. Hence it has high porosity or low porosity, but only rarely a 'middle' porosity. As expected, we see that the true curve switches back and forth between the extremes of the distribution but does not usually fall in the middle. The simulations, since they sample from this conditional distribution must have the same switching effect, hence there will be less 'shoulder effects' in channel intervals than with a continuous Gaussian model.

Finally, in figure 16 we compare the same well section for the case of the incorrect, short variogram for both the Ember model and the Gaussian case. Since the estimates of probability and the quantiles haven't changed, the only difference in the Ember model is the variation in simulations, which is a bit higher. On the contrary, the Gaussian case (calculated a posteriori from hundreds of simulations) shows that the probability estimates are too smooth. The simulations are too random, and the quantiles are far too wide, especially in the shoreface zones.

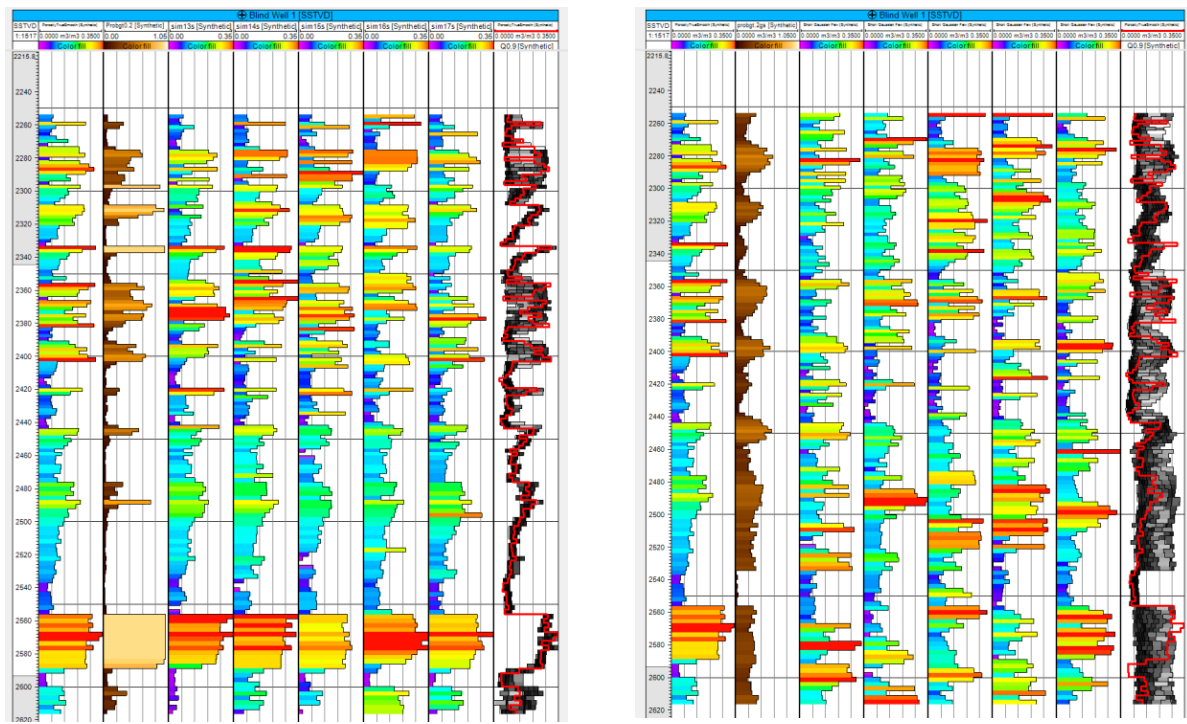

**Figure 16:** Same result as in figure 15, with short range variogram. Ember on left. Gaussian on right. Notice that the Gaussian envelope is almost uninformative compared to Ember.

# 5 Appendix

A Decision (or Random) Forest is an ensemble of Decision Trees. Decision Trees have been used for non-parametric regression for many years and their theory is reasonably well understood (Győrfi et al., 2002). They work by inducing a partition of space into hyper-rectangles with one or more data in each hyper-rectangle. To be concrete, suppose we have observations $(Z_i, Y_i), i = 1, ..., n$ with each $Y_i$ being a vector of predictor variables of dimension $p$ and $Z_i$ the $i^{th}$ training observation of a target variable that we wish to estimate. This paper is concerned with spatial data, so that we are looking for estimation or simulation of $Z(x)$ at a location $x$. The notation $Z_i$ is thus shorthand for $Z(x_i)$, the value of the target variable at the observed location $x_i$. The basic idea behind decision trees is to construct the tree using the training data $(Z_i, Y_i)$ and to predict at a new location x by passing the vector $Y(x)$ down the tree reading off the values of $Z_i$ stored in the terminal node hyperrectangle, using them to form the required type of estimator.

A decision tree is grown by starting with all data in the root node. Nodes are split reclusively using rules involving a choice about which of the $p$ coordinates/variables $Y^p$ in $Y$ to use as well as the value of that variable used for the split, $s_p$, with vectors $(Z_i, Y_i)$ going to the left child if $Y_i^p < s_p$, and to the right child if not. When tree growth terminates (another rule about size of terminal node), the geometry of the terminal node is therefore a hyper-rectangle. The associated set of $Z_i$ are stored in that terminal node. In random forests the choices of variables and splits are randomized in ways explained in the next paragraph, but for now let $\theta$ be the set of parameters chosen and call $T(\theta)$ the tree created with this choice of parameters. Denote by $R$ the hyper-rectangles associated with the leaves of the tree and call $R(y, \theta)$ the rectangle which represents the terminal node found by dropping $Y(x) = y$ down the tree $T(\theta)$. As stated before, estimations will be created from these terminal nodes, so for example, the estimate of the mean value of $Z(x)$ given $Y(x) = y$ from the single decision tree $T(\theta)$ is

$$\hat{\mu}(y, \theta) = \sum_{i \epsilon R(y,\theta)} \omega_i(y, \theta) Z_i$$

where the weight is defined by $\quad \omega_i(y, \theta) = \dfrac{\mathbb{I}_{\{Y_i \in R(y,\theta)\}}}{\sum_j \mathbb{I}_{\{Y_j \in R(y,\theta)\}}}$ i.e., the weight is 1 divided by the number of data in the hyper-rectangle when $Y_i$ is in the terminal node and is 0 when it's not.

Decision Forests are just ensembles of decision trees. The Random Forest is a specific type of decision forest introduced by Leo Breiman (1999, 2004) which have been particularly successful in applications combining the ideas of Bagging (Breiman, 1996) and Random Subspaces (Ho, 1998). In most versions of decision forest, randomization is performed by selecting the data for each tree by subsampling or bootstrapping and by randomizing the splitting. We extend the parameter $\theta$ to cover the data selection as well as splitting. An adaptive split makes use of the $Z_i$ values to help optimize the split. A popular choice is the CART criterion (Breiman et al., 1984), whereby a node $t$ is split to $t_l = \{i \in t; Y_i^p < s_p,\}$ and it's compliment $t_r$. The variable $p$ is chosen among a subset of the variables chosen randomly for each split such that the winning split variable and associated split value are the ones that minimize within node variance of the offspring. A slightly more radical split is employed in this paper. Instead of testing all possible split values for each variable, a split value is chosen at random for each variable candidate and then, as before, the within node variances for children calculated and the minimum chosen. This additional randomness seems to help reduce visual artefacts when using a random forest spatially.

The weight assigned to the *i*th training sample for a forest is $\omega_i(y) = \frac{1}{k}\sum_{i=1}^{k} \omega_i(y, \theta_i)$ where k is the number of trees. The estimator of the conditional distribution, which we need for Ember is then (Meinshausen, 2006)

$$\hat{F}(Z(x) = z|Y = y) = \sum_{i=1}^{n} \omega_i(y) \mathbb{I}_{\{Z_i < z\}}$$

And the forest estimate of the conditional mean is $\hat{\mu}(x|y) = \sum_i \omega_i(y) Z_i$.

There is a large literature on Random Forests, so we only mention a few that were helpful to the current work. As well as the foundational papers by Breiman and Meinshausen, Lin and Jeon (2006) describe how they can be interpreted as k-PNN, a generalization of nearest neighbour algorithms and give some good intuitive examples to help understand the role of weak and strong variables, splitting strategies and terminal node sizes. Establishing consistency is relatively easy for Random Forests with some reasonable conditions (we look at an example later). However, there has not of yet been a full explanation of how Breiman's Forest achieves such good results. Nonetheless, some good progress has been made, extending the basic consistency results. The PhD of Scornet (2015) includes a demonstration of the consistency of the Breiman forest in the special case of additive regression models (that chapter was in conjunction with Biau and Vert). Other results concern Central Limit Type results for Random Forests (Athey et al (2019) and references therein). Mentch and Zhou (2019) investigate the role of the regularization implicit in Random Forests to help explain their success.

## 5.1 Embedding Models in Decision Forests

If we have one or more models, $M_j(x), j = 1, \ldots, l$, which are themselves estimators of $Z(x)$, it is possible to embed these for use within a random forest. As described in the main article, the reason we might want to do this is to make use of information that is not directly available in the form of observed data. In spatial modelling, the variogram or covariance function can give information about the spatial continuity of the variable of interest. A model, such as kriging (Chiles and Delfiner, 2012), which is based on this continuity is well known to be a powerful technique for spatial estimation. In cases where there are a number of other predictor variables (often called secondary variables in the Geostatistics literature), the standard method of combining the information is through linear models of coregionalization (Wackernagel, 2003). The linearity of the relationship between variables can be restrictive and correlations may not be stationary. These considerations, as well as the possibility of a simplified workflow have prompted the idea of embedding models, which allow the Ember model to capture most of the power of the embedded model while allowing for better modelling of the relationship between variables and of local variability of the target variable. The embedding is *lazy* in the sense that rather than looking for potentially very complex interactions between variables, it tries to use the recognized power of Random Forests to deal with this aspect of the modelling. The goal here is produce an estimate of the conditional distribution $\hat{F}(Z|Y, M)$.

In the sort of applications considered here, the *Y* variable will usually include the spatial location, x, meaning that the forest will explicitly use spatial coordinates as training variables. This means that at each location, we can estimate the conditional distribution $\hat{F}_x \stackrel{\text{def}}{=} \hat{F}(Z(x)|Y(x) = y, M(x) = m)$, with *Y* a vector of size *p* and *M* a vector of size *l*. We call this family of estimates the envelope of Z. Note, this is not the full conditional distribution at x, rather it will be treated as a marginal distribution at x that we will sample from in appendix C. To emphasize that this is not a model of a spatial random function, we note that the mean of $\hat{F}_x$, $\hat{\mu}_x$, is a trend rather than an exact interpolator, although in Appendix B, we see that it generally performs well.

Training a forest with embedded models requires an extra step. In many cases the embedded model is an exact interpolator so $M(x_i) = Y_i$. This is the case if the embedded model is kriging. If training were to use this exact value, then the forest would over-train and learn little about how the model behaves away from the well locations, so this is not a useful strategy. A more promising strategy is to use cross validated estimates. Let $m_{-i}$ be the cross validated values found by applying the spatial model at data location $x_i$, using all the training data except that observed at $x_i$. This leads to the training algorithm

1) Choose the iid parameters $\theta_i$ for the ith tree.
2) Select a subset of data, the in-bag samples, for construction of the ith tree (either by subsampling or bootstrapping)
3) For each model and each in-bag sample, using all other in-bag samples, calculate the cross validated estimate $m_{-i}$.
4) Construct the ith tree using $(Z_i, Y_i, m_{-i})$

Since the model has now been 'converted into' a datum at each location $x_i$, to keep the notation simple, we will continue to talk about training with $(Z_i, Y_i)$ but need to remember that some of the $Y_i$ are actually embedded models. The effects of this are that

a) each tree is constructed on a different data set (as the embedded models depend on the actual in-bag sample)
b) Since the $m_{-i}$ are not independent, we can no longer make the iid assumption for $(Z_i, Y_i)$ that is usually made for random forests.

## 5.2 Consistency of the model

Meinshausen (2006) shows that his quantile Random Forest algorithm is consistent in that the estimate of the conditional distribution converges to the true conditional distribution as the number of samples increase.

The assumptions needed for theorem A.1 are

1) Y is uniform on $[0,1]^p$. In fact, it is only required that the density is bounded above and below by positive constants. We discuss what this means for embedded models later.

2) For node sizes,

   A) The proportion of observations for each node vanishes for large n (i.e. is o(n)).

   B) The minimum number of observations grows for large n (i.e., 1/min_obs is o(1)).

3) When a node is split, each variable is chosen with a probability that is bounded below by a positive constant. The split itself is done to ensure that each sub-node has a proportion of the data of the parent node. This proportion and the minimum probabilities are constant for each node.

4) $F(z|Y = y)$ is Lipschitz continuous.

5) $F(z|Y = y)$ is strictly monotonically increasing for each y.

**Theorem A.1 (Meinshausen)** When the assumptions above are fulfilled, and if, in addition, the observations are independent, the Embedded Forest estimate is consistent pointwise, for each y,

Theorem A.1 remains valid when using embedded variables provided the hypothesis hold and the samples can be considered iid. The variables are iid when $U(x) = F(Z(x)|Y(x))$ are iid. While this is not necessarily the case, it is often quite a good approximation and may justify an embedding approach with general models. The hypothesis is not required when kriging is a strong embedded variable. Theorem A.2 (Daly, 2020) is quite simple. It is known that as data becomes dense around a point to be estimated, that kriging converges to the true answer. This just shows that the same is true here when either the embedded variable is kriging and is a strong variable or when the spatial coordinates x,y,z are strong variables. For the demonstration, we need to strengthen assumption 1 and weaken assumption 4. Let $\mathcal{D}$ a closed bounded subset of $\mathbb{R}^n$.

1a) The embedded data variables in Y are uniform on $[0,1]^p$. In fact, it is only required that the density is bounded above and below by positive constants. Moreover the samples occur at spatial locations $S_n = \{x_i\}_{i=1}^n$ such that, for any $\epsilon > 0$, $\exists N$ such that $\forall x \in \mathcal{D}, \exists x_l \in S_n$ with $d(x, x_l) < \epsilon$ for all $n > N$. In other words, the samples become dense uniformly.

4a) $Z(x), x \in \mathcal{D}$ is a continuous mean squared second order random function with standard deviation $\sigma(x) < \infty$.

**Theorem A.2** Let $\mathcal{D}$ a closed bounded subset of $\mathbb{R}^n$. Let $Z: \mathcal{D} \to L^2(\Omega, \mathcal{A}, p)$ be a continuous mean squared second order random function with standard deviation $\sigma(x) < \infty$, and if, in addition, assumptions 1a, 2, 3 and 5 hold, then

1) If kriging is an embedded model and it is a strong variable for the Random Forest, then the Ember estimate $\hat{\mu}(x|y) = \sum_i \omega_i(y) Z_i$, converges in $L^2$ to Z(x).

2) If the random forest also has a mean, $m(x)$, that is continuous in x, is trained on the coordinate vector x, and if each component of x is a strong variable, then the Ember estimate $\hat{\mu}(x|y) = \sum_i \omega_i(y) Z_i$, converges in $L^2$ to Z(x).

For simulation, we need a full probability model. Defining the sampling variable as $U(x) = F(Z(x)|Y(x) = y)$, the required hypothesis is that the sampling variable is a uniform m.s continuous stationary ergodic random function. This means that the class of random functions that we can model are of type $Z(x) = \psi_x(U(x))$ for some monotonic function $\psi_x$, which depends on x. To model $Z(x)$ let us further assume that $U(x)$ is generated as the transform of a standard multigaussian random function. So $U(x) = G(X(x))$ with G the cumulative Gaussian distribution function and X(x) is a stationary Multigaussian RF with correlation function $\rho(h)$. This leads to the relation between Z and X,

$$Z(x) = \varphi_x(X(x)) \equiv \psi_x(G(X(x)))$$

This can be expanded in Hermite polynomials (Daly, 2020), which when locations is considered to be a random variable and the expansion is truncated to first order, give equation 3.

To obtain a conditional simulation, it is sufficient to choose a value of $u_i = U(x_i)$ at data location $x_i$ such that the observed target variable data $z_i$ is $z_i = F_{x_i}(u_i)$ where $F_{x_i}$ is the envelope distribution at $x_i$. Now in general there is more than one such possible $u_i$, particularly when the envelop distribution is estimated using random decision forests. If fact any $u \in [u_{low}, u_{high}]$ will satisfy this condition where $u_{low}, u_{high}$ can be read off the Ember estimate of $F_{x_i}$. Since in this section, we assume that u is created as a transform of a Gaussian RF, then a truncated Gaussian distribution can be used for the sampling (e.g. Freulon, 1992). The simulation process becomes,

1) Infer a covariance for the underlying Gaussian X(x) (for example, using equation 3)
2) For each simulation:
   a. Use a truncated gaussian to sample values of $u_i$ satisfying $z_i = F_{x_i}(u_i)$
   b. Construct a conditional Uniform Field U(x) based on X(x) such that $U(x_i) = u_i$
   c. Sample from $F(Z(x)|Y(x) = y)$ using U(x) for each target location x.

Note: The simulation is based on resampling the training data. So, if the sampling is close to uniform (i.e. each sample ends up in a hyperrectangle of approximately the same hypervolume), then the simulation will roughly follow the same distribution as the training data.